\def\year{2024}\relax
\documentclass[letterpaper]{article} % DO NOT CHANGE THIS
\usepackage{aaai22}  % DO NOT CHANGE THIS
\usepackage{times}  % DO NOT CHANGE THIS
\usepackage{helvet}  % DO NOT CHANGE THIS
\usepackage{courier}  % DO NOT CHANGE THIS
\usepackage[hyphens]{url}  % DO NOT CHANGE THIS
\usepackage{graphicx} % DO NOT CHANGE THIS
\usepackage{natbib}  % DO NOT CHANGE THIS AND DO NOT ADD ANY OPTIONS TO IT
\usepackage{caption} % DO NOT CHANGE THIS AND DO NOT ADD ANY OPTIONS TO IT
\usepackage{subcaption}
\DeclareCaptionStyle{ruled}{labelfont=normalfont,labelsep=colon,strut=off} % DO NOT CHANGE THIS
\usepackage{CJKutf8}

\usepackage{algorithm}
\usepackage{algorithmic}
\usepackage{balance}
\usepackage{newfloat}
\usepackage{listings}
\floatstyle{ruled}
\newfloat{listing}{tb}{lst}{}
\floatname{listing}{Listing}
\title{Tracing the Unseen: Uncovering Human Trafficking Patterns in Job Listings}
\author {Siyi Zhou, Richard Peng, Emilio Ferrara}
\title{Tracing the Unseen: Uncovering Human Trafficking Patterns in Job Listings}
\author {
    Siyi Zhou,\textsuperscript{\rm 1,3}
    Jiankun Peng, \textsuperscript{\rm 2,3}
    Emilio Ferrara \textsuperscript{\rm 1,2,3}
}
\affiliations {
    \textsuperscript{\rm 1} Annenberg School of Communication and Journalism, 
    \textsuperscript{\rm 2} Viterbi School of Engineering, 
    \textsuperscript{\rm 3} USC Information Sciences Institute \\ 
    University of Southern California\\
    zhousiyi@usc.edu, pengrich@usc.edu, emiliofe@usc.edu
}
\begin{document}

\maketitle

\begin{abstract}
% Human trafficking is a complex and multifaceted issue that has been exacerbated by the advent of social media and online platforms. While existing works have focused on detecting trafficking victims through modeling escort ads, we wish to prevent such crimes by finding out which users have high likelihood of being a trafficker. As a stepping stone to this goal, we present our collection of 258,619 job posts from eight major regions in the US, spanning from 2006 to 2024, in this paper. Then we conduct an exploratory analysis by categorizing them into nine groups based on recruiters' self-reported information. By examining the industry types, contact information, and posting frequency, we revealed a significant preference for phone contact across various industries and suggested how external events can potentially affect job posting activity. This study highlights the need for further investigation into the role of online job boards and contact methods in facilitating human trafficking and the importance of developing strategies to combat this issue.

In the shadow of the digital revolution, the insidious issue of human trafficking has found new breeding grounds within the realms of social media and online job boards. Previous research efforts have predominantly centered on identifying victims via the analysis of escort advertisements. However, our work shifts the focus towards enabling a proactive approach: pinpointing potential traffickers before they lure their preys through false job opportunities. 
In this study, we collect and analyze a vast dataset comprising over a quarter million job postings collected from eight relevant regions across the United States, spanning nearly two decades (2006-2024). 
The job boards we considered are specifically catered towards Chinese-speaking immigrants in the US. 
We classify the job posts into distinct groups based on the self-reported information of the posting user. Our investigation into the types of advertised opportunities, the modes of preferred contact, and the frequency of postings uncovers the patterns characterizing suspicious ads. Additionally, we highlight how external events such as health emergencies and conflicts appear to strongly correlate with increased volume of suspicious job posts: traffickers are more likely to prey upon vulnerable populations in times of crises. This research underscores the imperative for a deeper dive into how online job boards and communication platforms could be unwitting facilitators of human trafficking. More importantly, it calls for the urgent formulation of targeted strategies to dismantle these digital conduits of exploitation.

\end{abstract}

\section{Introduction}

\paragraph{}

The scourge of human trafficking presents a formidable challenge in measuring and combating its reach, often muddled by the overlapping definitions with human smuggling. This confusion was clarified in the early 2000s by the United Nations, drawing a clear line between these nefarious activities while acknowledging their frequent co-occurrence \cite{lg:2003,un:2003}. In the United States, the narrative is particularly poignant among immigrants, who form the majority of trafficking victims, regardless of their smuggling status upon entry \cite{aclu:2007}. Unraveling the journey of these individuals is crucial to understanding their victimization and identifying the perpetrators.

Traditionally, the nexus between traffickers and their victims was forged in person, within tightly-knit communities. However, the digital age, marked by the ascendancy of social media, has transformed this dynamic, facilitating connections between potential victims and traffickers across vast distances \cite{f:2016}. Platforms like \textsl{chineseinla.com} serve as a double-edged sword for the Chinese-speaking diaspora in Los Angeles, offering vital community support alongside avenues for illicit employment and exploitation.

Highlighting the urgency of this issue, the \textit{UNODC's 2023 Trafficking in Persons Report} reveals that a significant portion of trafficking victims between 2012 and 2022 were ensnared through online recruitment methods, with an even larger group being indirectly exploited via digital platforms, including dating apps \cite{un:2024}. The US Department of State further underscores the role of online job boards as hunting grounds for traffickers aiming at vulnerable populations \cite{dos:2024}.

Despite these alarming trends, current anti-trafficking strategies have largely centered on mitigating risks associated with dating apps and escort services, deploying tools such as free background checks, and tools to safeguard or identify victims online \cite{wjef:2018,tzjm:2017,zlj:2019, lvklpjrf:2021, tl:2024,k:2023,p:2023}. Yet, these initiatives stop short of addressing the root of the problem: preemptively identifying traffickers through the guise of legitimate employment offers.

This paper seeks to bridge this gap by turning the lens towards locally-targeted Chinese speaking platforms within the United States, often the first port of call for new immigrants. Through a detailed examination of dubious job advertisements on these platforms, we aim not only to shine a light on potential trafficking operations but also to pioneer detection methods capable of unmasking human traffickers lurking behind the darker corners of the Web.

\section{Methodology}

% \paragraph{Navigating culturally niched communities.}
% Existing research on human trafficking has focused on understanding the demographics of vulnerable populations to aid in preventing trafficking across specific regions, genders, and age groups. Kevin Bales analyzed a comprehensive dataset of human trafficking cases from various countries, utilizing variables from the World Statistics Pocketbook, to demonstrate that poverty and lack of opportunity are often significant predictors of human trafficking incidents. Building on this insight, we created hypothetical scenarios in which individuals might be targeted by human traffickers. We identified keywords commonly used by these populations and conducted manual searches for job advertisements on Google in multiple languages. Table 1 outlines the three scenarios we developed, along with the corresponding keywords (translated to English) used to investigate where these populations seek employment opportunities.Through this search, we found 8 domains niched in Chinese speaking communities in the US with the most active job posting activities. \\

\subsection{Navigating Culturally Niched Communities}

Our investigation delves into the intricacies of culturally specific communities, recognizing that the fabric of human trafficking is woven with threads of demographic vulnerabilities. Pioneering work by Kevin Bales \cite{b:2007}, which meticulously combed through a global dataset of trafficking cases alongside variables from the World Statistics Pocketbook, has illuminated the stark reality that poverty and limited economic opportunities are often harbingers of trafficking risks. Inspired by Bales' findings, our study crafts a series of hypothetical yet realistic scenarios to simulate potential targeting by traffickers within distinct demographics, emphasizing the nuanced interplay between socioeconomic factors and vulnerability to exploitation.

Next, we ventured into data collection, pinpointing specific lexicons resonating with at-risk individuals. Our methodology involved an extensive, multilingual Web search for job advertisements, meticulously combing through the digital landscape to uncover where vulnerable demographics might seek employment. The culmination of this effort is encapsulated in Table \ref{Tab:Tcr1}, which presents three examples of constructed scenarios along with a set of keywords--translated into English for clarity--aimed at piercing the veil of where these communities converge in their quest for work.

This strategic search unveiled a constellation of eight domains, each a hub within the Chinese-speaking diaspora in the United States, teeming with job postings. These domains, on the surface ripe with opportunities,  potentially mask the sinister undertones of trafficking under the guise of legitimate employment. Our approach not only highlights the importance of nuanced, culturally sensitive research in unveiling trafficking pathways, but also sets the stage for more targeted interventions within these specific communities.

\begin{table}[h]
    \centering
    \footnotesize
    \begin{tabular}{|p{4.5cm}|p{3cm}|} 
        \hline
        \textbf{Scenario} & \textbf{Keyword} \\ 
        \hline
        New immigrants from China with low English proficiency looking for a job & Well-paid, Chinese language job, flexible job  \\
        \hline
        Newly graduated international students looking for jobs that offer visa and green card sponsorship & Sponsor, H1B, EB-2, EB-3, OPT  \\
        \hline
        Chinese speaking populations living in the US without legal work authorizations looking for under the table jobs that pay cash & Cash, under the table, temporary work, part time \\
        \hline
    \end{tabular}
    \caption{Example scenarios and corresponding keywords}
    \label{Tab:Tcr1}
\end{table}

\begin{figure}[t]
\centering
\includegraphics[width=\columnwidth]{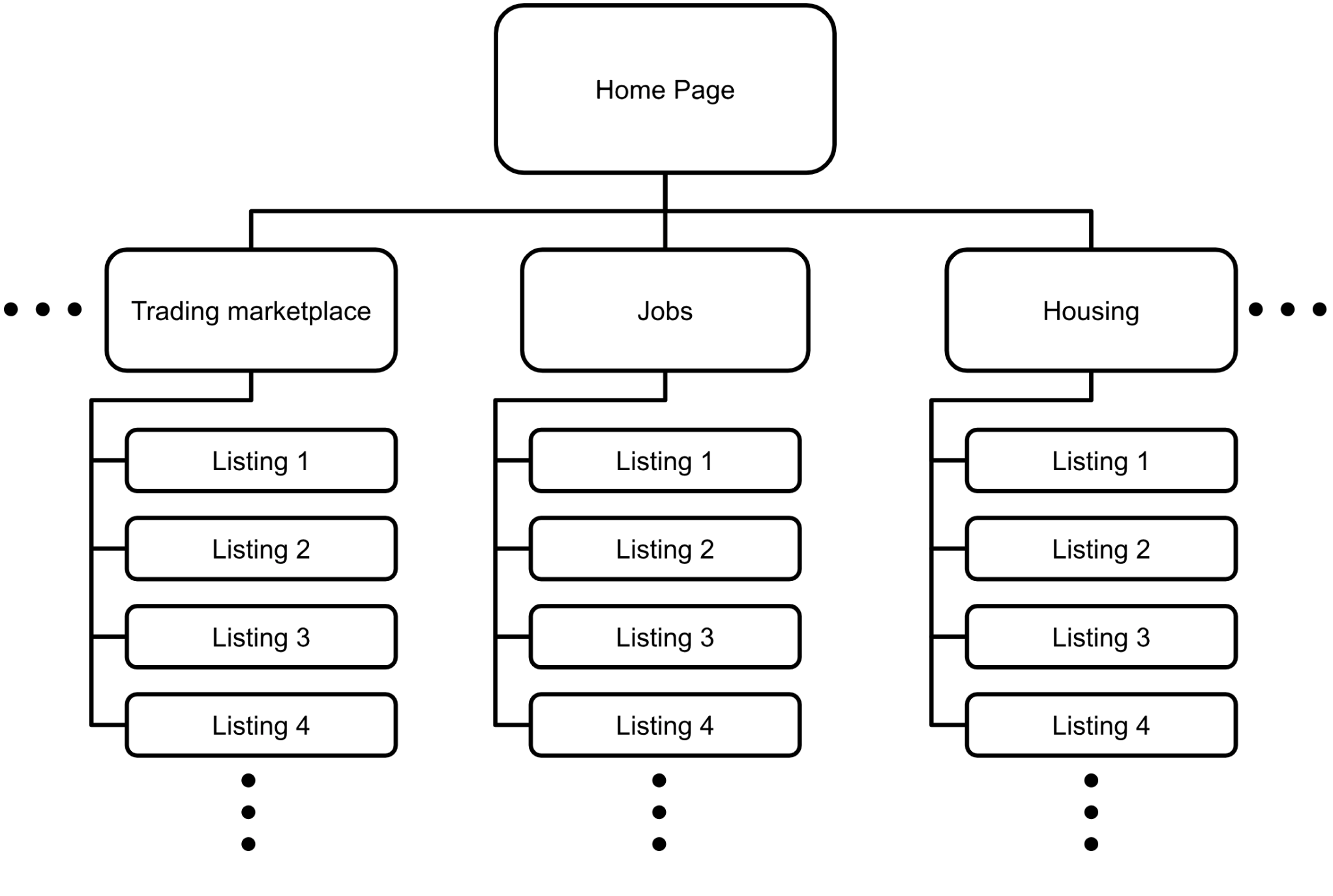} % Reduce the figure size so that it is slightly narrower than the column. Don't use precise values for figure width.This setup will avoid overfull boxes.
\caption{A simplified web structure of job ads platforms }
\label{fig1}
\end{figure}

% \paragraph{Pipelining ads collection across different platforms.}
% Our analysis of various domains revealed a consistent web structure, enabling the implementation of a streamlined collection process. Figure 1 illustrates the generalized structure of these domains. We employ a breadth-first search approach based on this structure. Initially, we gather all the links from the job listings section. Subsequently, we retrieve the source code of each link and store the HTML snapshot locally. This step ensures continued access to the data points, even if some pages are later removed. After accumulating all the HTML snapshots, we parse and convert them into the desired JSON format for further analysis.\\

\subsection{Job Ads Collection Across Different Platforms}

In dissecting the digital anatomy of various domains central to our investigation, a serendipitous discovery was made: a uniformity in web architecture across these platforms. This uniformity not only piqued our interest but also paved the way for a more efficient method of data collection. Figure \ref{fig1}, a schematic representation, sheds light on the commonalities, serving as the blueprint for our data scraping approach.

Leveraging the predictability of these structures, we adopted a breadth-first search strategy as our modus operandi. The initial phase involves a comprehensive sweep of the job listings section, aggregating every accessible link. Following this, we delve deeper, extracting the source code from each link and archiving it as an HTML snapshot on our local storage system. This preemptive measure safeguards against the potential loss of data, ensuring its availability for subsequent analysis even in the event of page removal or content alteration.

The final step of our collection sees the transformation of these HTMLs into structured JSON files, ready for the analytical deep dive. This conversion process not only facilitates a streamlined analysis but also underscores our commitment to preserving the integrity and accessibility of the data collected from these digital platforms.

% \paragraph{Framing information from raw data.}
% After collecting snapshots of all the job advertisement posts, we analyze the webpages and organize the data into columns. Each post is uniquely identified by its original URL as the key attribute. For each post, we extract details such as the type of hiring, contact information, job description, listing time, views, and authorship. These elements are valuable in several ways:
% 1) Type of Hiring: Helps categorize industries and identify which categories have more suspicious listings; 2) Contact Information: Useful for exploring the cross-domain network to determine where else this user or party has been active; 3) Job Descriptions: Contributes to existing language models that detect human trafficking, potentially offering more insights into human traffickers than victims.
% An example of the collected data is presented in Figure 2.\\

% \begin{figure}[t]
% \centering
% \includegraphics[width=\columnwidth]{sample.png} % Reduce the figure size so that it is slightly narrower than the column. Don't use precise values for figure width.This setup will avoid overfull boxes.
% \caption{An example of the data collected from one job ad }
% \label{fig2}
% \end{figure}

\subsection{Extracting Information from Raw Job Ads Data}

The culmination of our data collection effort—snapshots of myriad job advertisements—marks the beginning of a critical phase: distillation and organization of data gleaned from these webpages. Central to our analytical framework is the assignment of a unique identifier to each post, with the original URL serving as the pivotal attribute. This step underpins the extraction of several key details from each advertisement, including but not limited to the type of hiring, contact information, job description, listing time, number of views, and authorship information.

The significance of these data points cannot be overstated, as they collectively offer a multifaceted lens through which the complex landscape of online job postings can be scrutinized:

\begin{enumerate}

\item Type of Hiring: This is instrumental in segmenting the data across different industries, enabling us to pinpoint sectors that might be more prone to suspicious activities.
\item Contact Information: A gateway to unraveling the cross-domain presence of users with trafficking intents, this information aids in mapping out the digital footprint of parties potentially involved in nefarious activities. 
\item Job Descriptions: A cornerstone for enhancing existing language models designed to detect signals of human trafficking. This information provides a deeper understanding of the modus operandi of traffickers, potentially offering novel insights into their tactics and strategies over their victims.

\end{enumerate}

\begin{CJK*}{UTF8}{gbsn}

\begin{table*}[t]
\centering \footnotesize
\begin{tabular}{|l|l|}
\hline
\textbf{Field} & \textbf{Value} \\
\hline
URL & \url{https://www.chineseinla.com/f/page_viewtopic/t_1962623.html} \\
\hline
Post Type & hiring \\
\hline
Hiring Type & internship \\
\hline
Industry & computer \\
\hline
Location & 1455 Monterey Pass Rd \#206 Monterey Park CA \\
\hline
Company Name & LaneCert Education Group \\
\hline
Title & \begin{tabular}[c]{@{}l@{}}优质转让,高新就业,小白也能轻松落地,留学生 OPT, H1B 高新落地\\ \end{tabular} \\
\hline
Email & info@lanecert.com \\
\hline
Phone & 3234880011, 6267265138, 6297265138 \\
\hline
Post & \begin{tabular}[c]{@{}l@{}}诚意出让七年业界口碑,低收WIOA \& 退役军人 VA GI BILL \\ 免费学CCIE 顶级讲师,一对一辅导,自有ISP实验室... \end{tabular} \\
\hline
Date Created & 2020/09/17 \\
\hline
Date Modified & 2024/01/21 \\
\hline
Total Number of Edits & 38 \\
\hline
Views & 57,020 \\
\hline
Author & LaneCert\_LA \\
\hline
Author Profile & \url{https://www.chineseinla.com/user/id_2385433.html} \\
\hline
\end{tabular}
\label{tab2}

\caption{An example of \textit{Job Posting} metadata information}
\label{table:job_posting}
\end{table*}
\end{CJK*}

Table \ref{table:job_posting} presents a snapshot of the organized data, showing the transformation of raw data into a structured repository ready for in-depth analysis. This process not only lays the groundwork for subsequent investigative endeavors but also enriches the toolkit available to researchers and practitioners fighting against the scourge of human trafficking.

\section{Content of our Job Ads Dataset}

This segment of our study delves into a detailed exploration and profiling of the dataset at hand. Although Table \ref{table:job_posting} exemplifies an ideal scenario—where every job listing is accompanied by a complete set of attributes—reality often tells a different story. A significant portion of the listings we encountered were bereft of one or more critical attributes, such as the industry type, hiring type, or geographical location. Our investigation, at this point, is primarily focused on unraveling the patterns associated with the methods of contact delineated within these listings, guided by the industry attribute as self-reported by the recruiters.

Moreover, our initial analysis extends to scrutinizing the temporal dynamics of job postings—how the frequency of listings by recruiters fluctuates over time across different industries. This twofold examination, encompassing both the modes of communication and the time series of posting activity, sheds light on the underlying correlations between online job advertisements and their potential role in facilitating human trafficking.

Through this analytical endeavor, we aspire to uncover the latent threads that may tie seemingly innocuous job postings to the dark underbelly of trafficking operations, thereby contributing to a more nuanced understanding of how digital platforms can be exploited for such heinous purposes.

% \section{Content}
% This section provides a general profile and an exploratory analysis of the data. While Figure 2 presents an ideal case where all attributes exist for the listing, most listings are missing at least one or two attributes, such as industry type, hiring type, or location. At this stage, we primarily focus on exploring the pattern of contact methods presented in the listings for different industries, using the recruiter's (i.e., user posting the ad) self-reported industry attribute. Additionally, we examine the posting frequency of recruiters over time for different industries. By analyzing their contact methods and posting frequencies, we aim to demonstrate the connections between these job posts and potential trafficking risks.\\

\subsection{A General Outline of the Job Ads Dataset}

In an effort to pierce through the veil of online job postings within the Chinese-speaking diaspora in the United States, our collection spans a vast array of 258,619 job advertisements. These postings, collected from niche platforms across eight major U.S. regions, trace back from the year 2006 up to 2024, offering a longitudinal view of the online job market's evolution. Table \ref{Tab:Tcr2} summarizes the distribution of these advertisements, laying out a diverse landscape of sources.

A  characteristic of our dataset is the uniform presence of certain attributes across all postings—namely, the title, author, job description, and creation date. However, the inclusion of other attributes remains variable, painting a picture of data richness contrasted by gaps in informational completeness. Further distinguishing these job posts is their certification status as indicated by the job post's address prefix: "j\_" signifies a platform-certified posting, while "t\_" denotes its uncertified counterpart. Our analytical focus sharpens on the "t\_" category, navigating through the uncertified postings to unearth patterns and insights.

\begin{figure}[h]
\centering
\includegraphics[width=\columnwidth]{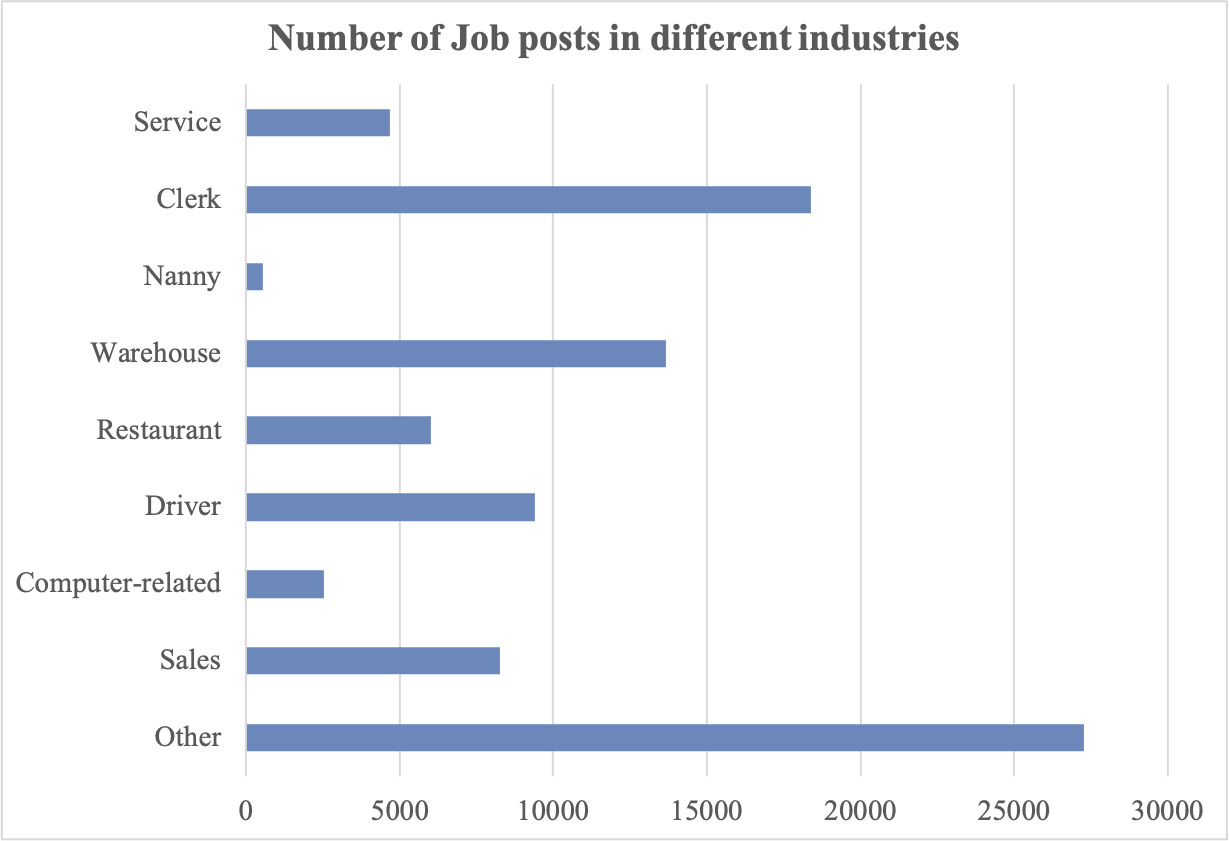} % Reduce the figure size so that it is slightly narrower than the column. Don't use precise values for figure width.This setup will avoid overfull boxes.
\caption{Number of job ads listed for different industries (self-reported by the posting users). The count does not include those listings that does not specify their industries }
\label{fig2}
\end{figure}

The dataset was methodically categorized into nine distinct job groups based on the recruiters' self-disclosed information, as shown in Figure \ref{fig2}. The findings reveal a significant demand for clerical (18,369), warehouse (13,673), and driving (9,406) positions, with the clerical category encompassing a wide array of roles including secretarial jobs. A deeper dive into this category—specifically, a random examination of 15 secretary job listings—uncovered a disconcerting pattern: some listing solicited interviews at a hotel, and its provided contact number was previously listed on an escort service website, signaling a glaring sex trafficking hazard. Moreover, an analysis of the warehouse and driver job postings revealed 1,294 ads with job descriptions shorter than 200 words. This brevity, marked by a lack of detail, potentially flags an increased risk of labor trafficking, underlining the intricate challenges and dangers lurking within these online job platforms.

\begin{table}[t]
    \centering
    \footnotesize
    \begin{tabular}{@{}l@{}c@{}cc@{}} 
        \hline
        Domain&Number of posts&Earliest&Latest\\
        \hline
        chineseinatlanta.com&3,142&2016-01-22&2024-03-12\\
        chineseinflorida.com&2,741&2015-12-24&2024-03-12\\
        chineseinla.com&142,325&2006-09-06&2024-03-12\\
        dcchinaren.com&2,580&2013-03-21&2024-03-12\\
        seattlechinaren.com&6,913&2013-03-17&2024-03-12\\
        vegaschinaren.com&3,588&2010-12-13&2024-03-12\\
        nychinaren.com&58,880&2011-05-13&2024-03-13\\
        chineseinsfbay.com&38,450&2011-01-11&2024-03-14\\
\hline
    \end{tabular}
    \caption{Composition of sources for our \textit{Job Ads} dataset}
        \label{Tab:Tcr2}
\end{table}

% \subsection{A general profile of the dataset}
% We collected a total of 258,619 job posts from niche Chinese-speaking platforms located in 8 major regions in the US, spanning from as early as 2006 to 2024. Table 2 provide a detailed profile of the distribution of these job posts across different platforms. All of them contain attributes of title, author, job description, and creation date, while the presence of other attributes is not guaranteed. The job posts can also be distinguished based on whether they are certified by the platform through the key - the address of the job post. "j\_" indicates that the job post is certified, and "t\_" indicates that it is not. Our framing process primarily focused on and adjusted for "t\_" job posts, the uncertified ones. \\

% Using the recruiters' self-reported information, we categorized the job ads into nine groups. Among these, clerk (18,369), warehouse (13,673), and driver (9,406) jobs were the most in-demand. The clerk category is broad and includes various jobs, such as secretary positions. We randomly applied 15 listings for secretary positions and found one that requested an interview at a hotel; the provided phone number also appeared on an escort website, indicating a potential risk of sex trafficking. Additionally, 1,294 posts in the warehouse and driver categories contained fewer than 200 words in their job descriptions. The lack of detail in these descriptions contributes to the risk of labor trafficking.\\

\subsection{Contact Methods by Industry}

In our exploration of job advertisement platforms, a striking pattern emerged regarding the preferred methods of contact across various industries. 

\begin{figure}[t]
\centering
\includegraphics[width=\columnwidth]{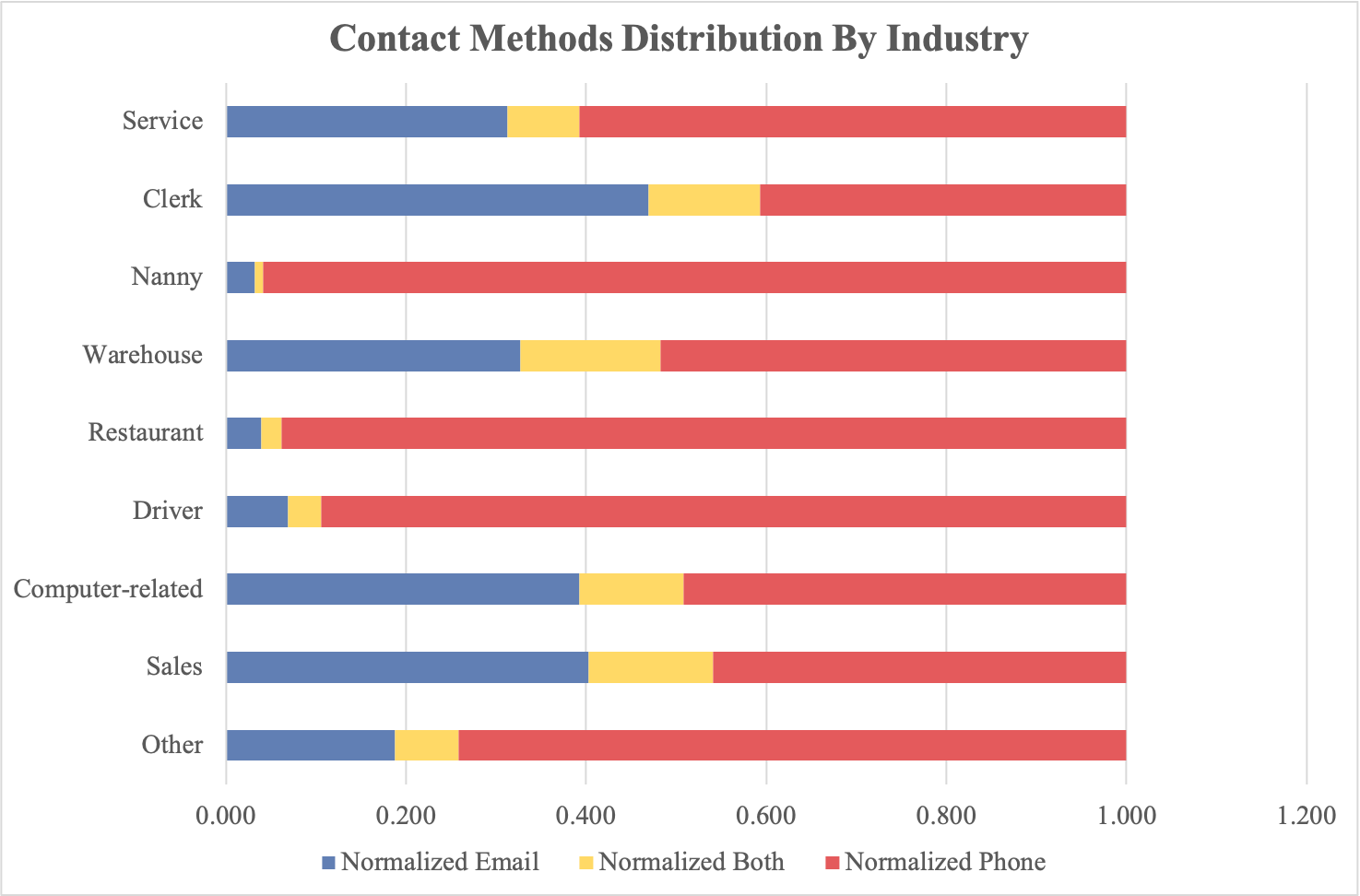} % Reduce the figure size so that it is slightly narrower than the column. Don't use precise values for figure width.This setup will avoid overfull boxes.
\caption{Percentage of preferred contact methods for different industries }
\label{fig3}
\end{figure}

Looking at Figure \ref{fig3}, it appears that an overwhelming 58.8\% of job postings favored telephone numbers as the primary mode of communication, with 30.6\% opting for email as their exclusive contact method. A smaller segment, 10.6\%, offered the choice of both phone and email, ensuring a broader avenue for potential applicants to reach out. Phone communication is harder to monitor and often leads to in person contacts, amplifying the risk of harm \cite{k:2023}.

A deeper dive into the data reveals an industry-specific trend, particularly within the massage parlor sector, where phone numbers are universally adopted as the sole contact method. This practice is not isolated; beyond the realms of clerical and computer-related jobs, a distinct preference for phone communication pervades, spanning sectors like nanny services, restaurants, and driving. These fields demonstrate a marked predilection for phone contacts, coupled with a discernible hesitancy to diversify communication channels.

Table \ref{Tab:Tcr3} elucidates this trend further, detailing the distribution of phone numbers and email addresses across the board for job applications. The reliance on phone numbers introduces a layer of anonymity and diminishes traceability, in stark contrast to the digital footprint left by email correspondences. This distinction has significant implications for the legitimacy of job advertisements and, by extension, the potential for exploitation. The preference for phone communication, while perhaps convenient, concurrently amplifies the vulnerabilities and risks associated with human trafficking, underscoring the need for a critical assessment of contact methodologies in job listings.

% \subsection{Contact methods by industry}
% We found that 58.8 percent of the job ads on these platforms list phone numbers as the sole method of contact, while 30.6 percent use email exclusively, and 10.6 percent provide both phone numbers and email addresses for contact. Notably, all job postings in the massage parlor category exclusively use phone numbers for contact.\\

% Beyond the clerk and computer-related sectors, a notable preference for phone contact is observed across various industries. Sectors such as nanny services, restaurants, and driving exhibit not only a strong inclination towards phone communication but also a notable reluctance to offer multiple contact methods. Table 3 presents a comprehensive breakdown of the phone numbers and emails provided for job applications. The use of phone numbers as contact information in job ads carries a higher degree of anonymity and lower traceability compared to emails. Consequently, this practice reduces the credibility of the job ads and elevates the risks associated with human trafficking.\\

\begin{table}[t]
\centering
\footnotesize
\begin{tabular}{lcccc}
\hline
Industry&Email&Phone&Both&total\\
\hline
Other&3,028&11,934&1,130&16,092\\
Sales&2,768&3,153&949&6,870\\
Computer-related&800&1,003&238&2,041\\
Driver&393&5,125&216&5,734\\
Restaurant&100&2,392&57&2,549\\
Warehouse&4,178&6,629&1,992&12,799\\
Nanny&11&325&3&339\\
Clerk&9,278&8,059&2,447&19,784\\
Service&900&1,748&228&2,876\\
\hline
\end{tabular}
\caption{Contact method preferences for different industries}
\label{Tab:Tcr3}
\end{table}

% \begin{table}[h]
%     \centering
%     \footnotesize
%     \label{Tab:Tcr1}
%     \begin{tabular}{lcccc} 
%         \hline
%         Industry&Email&Phone&Both&total\\
%         \hline
%         Other&3028&11934&1130&16092\\
%         Sales&2768&3153&949&6870\\
%         Computer-related&800&1003&238&2041\\
%         Driver&393&5125&216&5734\\
%         Restaurant&100&2392&57&2549\\
%         Warehouse&4178&6629&1992&12799\\
%         Nanny&11&325&3&339\\
%         Clerk&9278&8059&2447&19784\\
%         Service&900&1748&228&2876\\
%         \hline
%     \end{tabular}
%     \caption{Contact method preferences for different industries}
% \end{table}

\begin{figure*}[h]
\centering
\includegraphics[clip, trim=45 45 50 45, width=2.1\columnwidth]{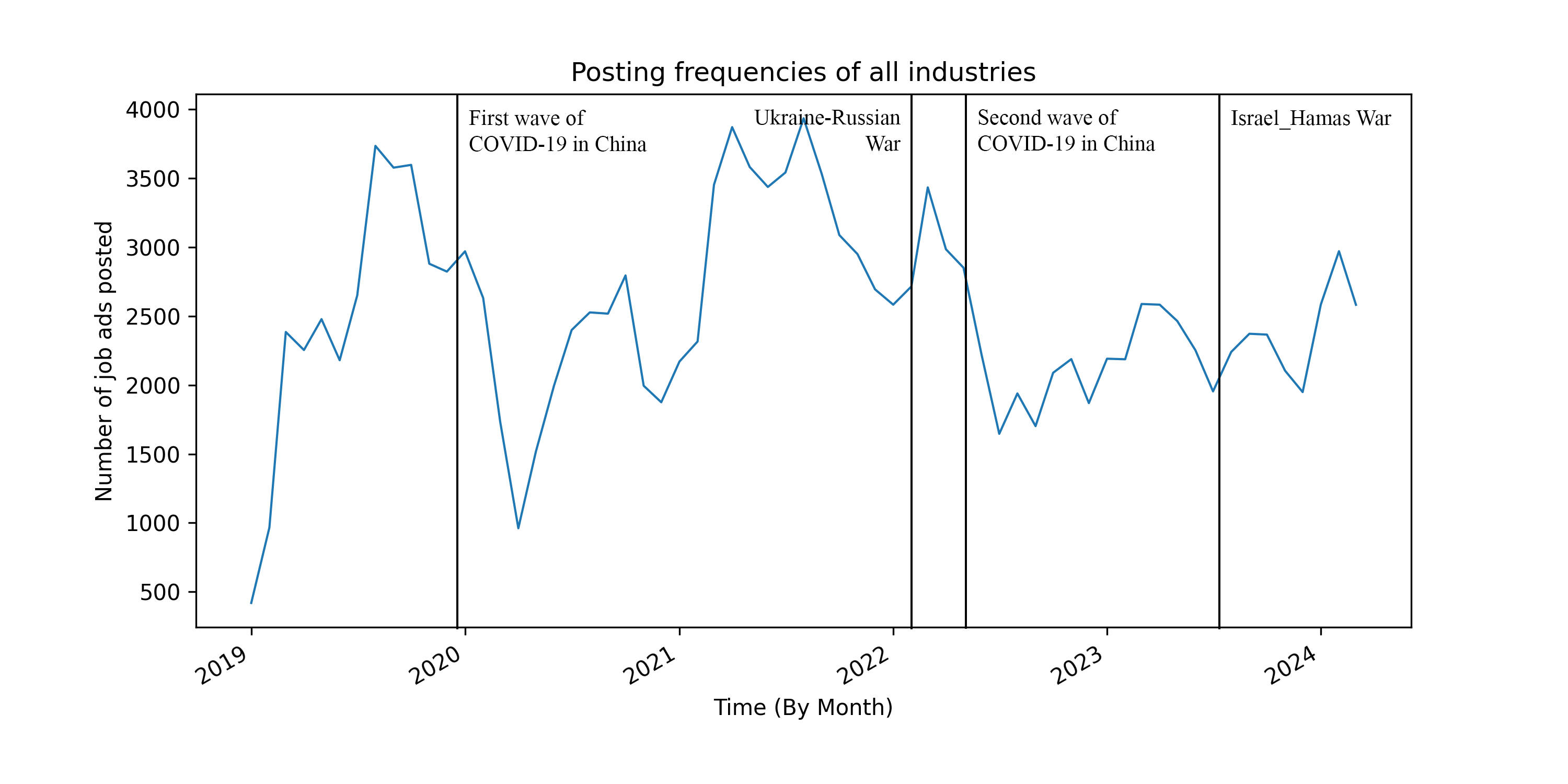} % Reduce the figure size so that it is slightly narrower than the column. Don't use precise values for figure width. This setup will avoid overfull boxes.
\caption{Posting frequencies of all job posts from 2019 to 2024}
\label{fig4}
\end{figure*}

\begin{figure*}[h]
\centering
\includegraphics[clip, trim=45 45 50 45, width=2.1\columnwidth]{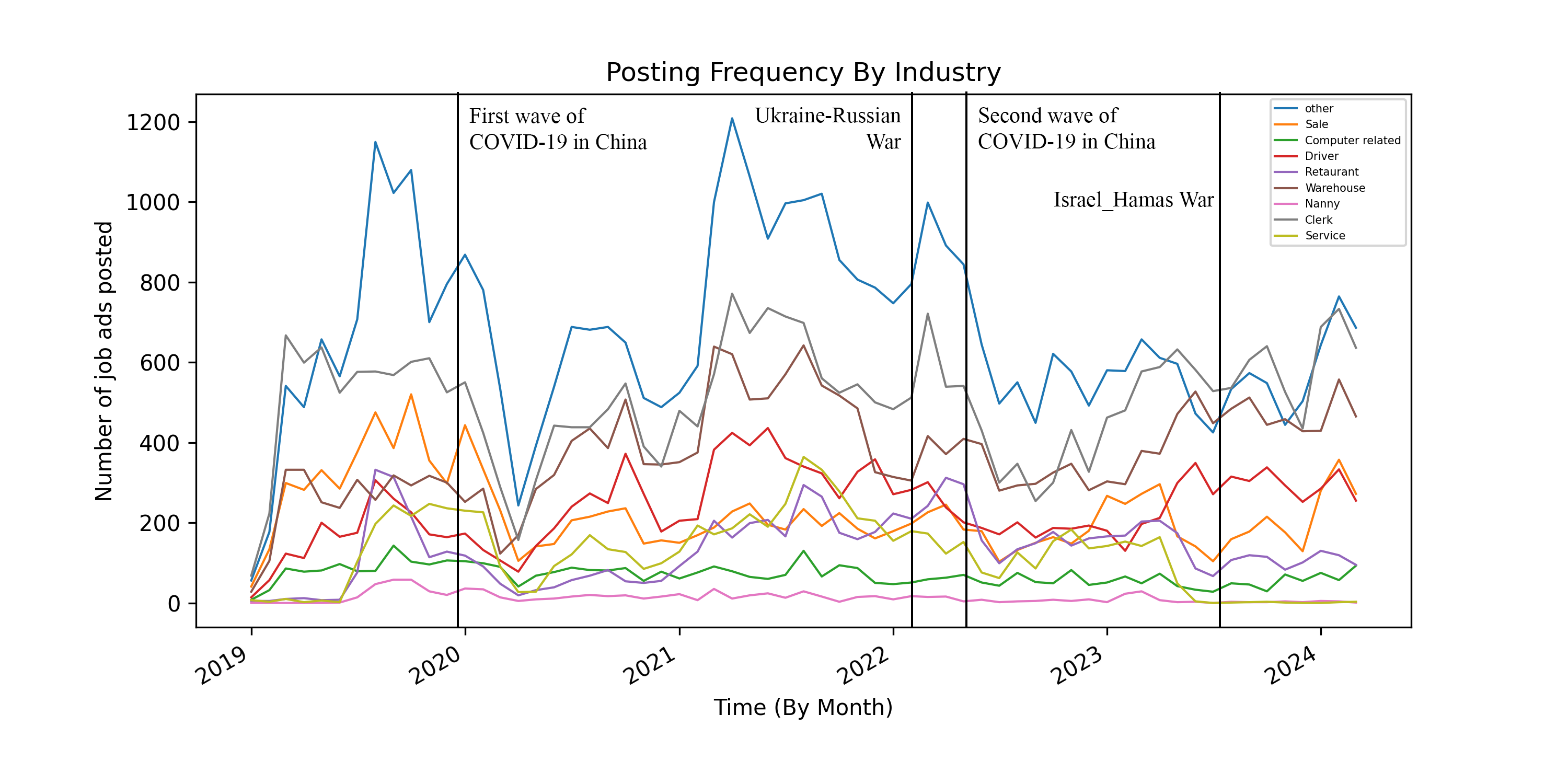} % Reduce the figure size so that it is slightly narrower than the column. Don't use precise values for figure width. This setup will avoid overfull boxes.
\caption{Posting frequencies for jobs ads in different industries from 2019 to 2024}
\label{fig5}
\end{figure*}

\subsection{Job Ads Posting Frequency Over Time}

Our next analysis, depicted in Figures \ref{fig4} and \ref{fig5}, closely examines the temporal dynamics of job posting frequencies, both overall and within specific industries, against the backdrop of major global events over the past five years. Intriguingly, the data suggest that geopolitical tensions, such as the Ukraine-Russia conflict and the conflict between Israel and Hamas, had moderate effects on the volume of job advertisements. However, the impact of the COVID-19 pandemic, particularly within China, paints a different picture, significantly influencing recruiter behavior.

The timeline of the pandemic reveals two distinct periods of activity: the initial outbreak between January and June 2020 saw a sharp decline in job postings, mirroring the global uncertainty and lockdown measures of the period. A similar downturn was observed during the second wave, spanning April to December 2022. Conversely, 2021 emerged as a period of recovery and heightened activity, potentially signaling an increased eagerness among the population to seek employment opportunities in the U.S., possibly driven by desires for immigration or better prospects amid the pandemic's economic fallout.

While the overarching trends in posting frequency appear consistent across industries, a closer inspection reveals nuanced variations. Industries such as nanny services and computer-related sectors showed remarkable resilience, remaining relatively unfazed by the highlighted events. On the other hand, the service, clerical, driving, and warehouse sectors demonstrated more pronounced fluctuations, indicating a higher degree of sensitivity to external disruptions.

This disparity in response rates across industries underscores the necessity for a deeper dive into the data. It prompts us to consider whether the vulnerability of certain job sectors to external events could serve as a barometer for detecting and understanding the behavior of human traffickers online. Such insights could be invaluable in devising more targeted strategies for combating trafficking in the digital age.

% \subsection{Job ads posting frequency over time}
% Figures \ref{fig4} and \ref{fig5} illustrate the posting frequency by recruiters overall and by industry, marked with selected major events that occurred in the past five years. Although the Ukraine-Russia War and the conflict between Israel and Hamas appear to have little impact on the number of posts, the prevalence and regulations of COVID-19 in China seem to have influenced recruiters' posting behavior. The first wave of the pandemic in China occurred between January and June 2020, during which we observed a steep drop in job posts. The second wave took place from April to December 2022, which also resulted in a decline in postings. In contrast, during 2021, there was a surge in job posts, which could be interpreted as an increased desire for immigration, stimulating job hunting in the US for this specific population.\\

% Although the patterns of job posting behavior are similar across all industries, we can see that nanny services and computer-related industries are least affected by these events. The posting behaviors of service, clerk, driving, and warehouse industries are more influenced by these events, given that they exhibit larger discrepancies in different months compared to nanny services and computer-related jobs. This observation warrants further investigation to determine whether the susceptibility of jobs in a certain industry can be used to predict and measure human traffickers' behavior online.\\

\section{Discussion and Potential Applications}

The collection of our Job Ads dataset marks a significant stride towards enriching the arsenal against online human trafficking, offering new angles to examine and thwart the intricate web of traffickers. Unlike conventional datasets that pivot around escort advertisements \cite{szekely2015building, kejriwal2017information, kejriwal2017knowledge, hundman2018always}, with a primary aim of rescuing victims post-factum, our dataset ventures into preemptive territory. It aspires not only to spotlight the perpetrators behind these heinous acts but also to stifle the propagation of their nefarious messages. Beyond its immediate utility in combating trafficking, this dataset might enable the study of immigration patterns and facilitate research within Asian American communities.

A pivotal insight from our exploratory analysis underlines a stark contrast in communication preferences on these platforms compared to more formalized channels like LinkedIn and Indeed. The predilection for phone communication in job advertisements, as revealed, is fraught with vulnerabilities. Phone numbers, with their cloak of anonymity and disposability, emerge as a double-edged sword. They afford traffickers the agility to swiftly weave a rapport with unsuspecting individuals, urging decisions under duress, away from the scrutinizing eyes of law enforcement. Despite the inherent challenges in tracing phone-based exchanges, these contact numbers inadvertently weave a thread that could unravel clandestine networks, potentially bridging the gap to escort services and enabling cross-platform hunts for trafficker identities \cite{ferrara2014detecting, catanese2012forensic}.

Moreover, the temporal trends captured in job postings shed light on the fluid dynamics of niche job markets and their susceptibility to human trafficking, especially within minimally regulated sectors. When dovetailed with datasets chronicling pivotal events or shifts in public policy and discourse, these trends hold promise for forecasting trafficker behavior with unprecedented precision.

Our methodological backbone is the industry classification as self-reported by the job posters, which, despite covering only a fraction (26.7\%) of the total dataset, significantly surpasses the NIH's recommended threshold for epidemiological studies \cite{mgbbd:2014}. This foundation not only legitimizes the robustness of our findings but also hints at the untapped potential of the unclassified posts, which, through linguistic analysis, could further refine our understanding of job market patterns and trafficking risks.

However, the dataset navigates through a sea of challenges. Temporal shifts in website content and inherent inconsistencies in job post attributes introduce a layer of complexity. While the universal presence of titles and job descriptions offers a lifeline for attribute extrapolation, the exclusion of deleted or moderator-removed postings casts a shadow on the dataset's comprehensiveness. This omission might skew the portrayal of the job market's darker underbellies, potentially glossing over the prevalence and markers of exploitative offers. Nonetheless, the sheer volume of data captured in real-time presents a compelling snapshot of the job market's current state, offering valuable insights for the ongoing battle against human trafficking.

\section{Conclusions}

In this study, we compiled a dataset of 258,619 job ads from Chinese-speaking online platforms in the US, covering two decades. This dataset offers a fresh perspective on combating online human trafficking, moving beyond the focus on escort ads. Our analysis highlights a widespread use of phone communication in job ads, raising concerns due to the anonymity and untraceability of phone numbers and their potential link to trafficking activities.

We also observed how global events like the COVID-19 pandemic affected job posting patterns, revealing varying degrees of industry resilience to external shocks. Despite facing limitations such as inconsistent data and missing posts, the comprehensive nature and focus on active ads provide significant insights into the job market's role in human trafficking, suggesting directions for future research and prevention strategies.

\paragraph{Data availability:}
Open data sharing has revolutionized computational social science research by fostering collaboration, enhancing the quality of research, and enabling the application of advanced computational methods \cite{davis2016osome, chen2021election2020, chen2023tweets}. Data access democratization allows researchers to delve into human behavior, social trends, and economic patterns more deeply and accurately and at scale \cite{chen2022charting, benjamin2023hybrid}. It supports interdisciplinary studies, combining fields like sociology and computer science, to tackle complex societal issues affecting vulnerable groups \cite{rao2021political, zhou2023unveiling}.  Making diverse datasets widely available also enables researchers to apply machine learning and data mining techniques to identify societal threats rapidly and at scale \cite{sapienza2018discover, tavabi2019characterizing, luceri2024unmasking}. 

In that spirit, we decided to make our Job Ads dataset openly available to the research community. We will also continue to maintain and expand it in the future.
Looking ahead, we hope other scholars will leverage our Job Ads dataset, alongside other available data, to dive into the complexities of human trafficking within the digital age, to not only enrich our comprehension of this global malaise but also to forge robust, data-informed frameworks for prevention and intervention.

% \section{Conclusions}

% In this paper, we present 258,619 job ads from niche Chinese-speaking platforms located in 8 major regions in the US posted in the past two decades. This dataset presents a novel approach to combating online human trafficking by providing an alternative perspective to existing data that primarily focuses on escort ads. Our analysis revealed a significant preference for phone contact across various industries. This preference for phone contact, coupled with the anonymity and low traceability of phone numbers, raises concerns about the credibility of job ads and the potential risks associated with human trafficking. Additionally, we examined the posting frequency of recruiters over time, noting the impact of major events such as the COVID-19 pandemic on posting behavior. Our findings suggest that certain industries, such as nanny services and computer-related sectors, are less affected by these events, while others exhibit larger discrepancies in posting frequency, potentially indicating susceptibility to human trafficking. While the dataset has limitations, such as inconsistent attributes and the exclusion of removed job posts, its large volume and focus on active posts make it a valuable resource for understanding the current state of the job market and developing strategies to combat human trafficking. Future research should continue to explore the potential of our Job Ads dataset and other alternative data sources to enhance our understanding of human trafficking and develop effective interventions.

\paragraph{Dataset:} 

https://github.com/serenalyoko/humantraffcking

\paragraph{Acknowledgements.} The authors are grateful to the Many Hopes foundation, a 501(c)3 organization that supports research to combat human trafficking.

% \newpage
\balance
\bibliography{paper}

\begin{thebibliography}{31}
\providecommand{\natexlab}[1]{#1}

\bibitem[{ACLU(2007)}]{aclu:2007}
ACLU. 2007.
\newblock Human Trafficking: Modern Enslavement of Immigrant Women in the United States.
\newblock \emph{American Civil Liberties Union}.

\bibitem[{Bales(2007)}]{b:2007}
Bales, K. 2007.
\newblock What Predicts Human Trafficking?
\newblock \emph{International Journal of Comparative and Applied Criminal Justice}, 31(2): 269--279.

\bibitem[{Benjamin et~al.(2023)Benjamin, Morstatter, Abbas, Abeliuk, Atanasov, Bennett, Beger, Birari, Budescu, Catasta et~al.}]{benjamin2023hybrid}
Benjamin, D.~M.; Morstatter, F.; Abbas, A.~E.; Abeliuk, A.; Atanasov, P.; Bennett, S.; Beger, A.; Birari, S.; Budescu, D.~V.; Catasta, M.; et~al. 2023.
\newblock Hybrid Forecasting of Geopolitical Events.
\newblock \emph{AI Magazine}.

\bibitem[{Catanese, Ferrara, and Fiumara(2012)}]{catanese2012forensic}
Catanese, S.; Ferrara, E.; and Fiumara, G. 2012.
\newblock Forensic analysis of phone call networks.
\newblock \emph{Social Network Analysis and Mining}, 3(1): 15--33.

\bibitem[{Chen, Deb, and Ferrara(2021)}]{chen2021election2020}
Chen, E.; Deb, A.; and Ferrara, E. 2021.
\newblock \#Election2020: The First Public Twitter Dataset on the 2020 US Presidential Election.
\newblock \emph{Journal of Computational Social Science}, 5: 1--18.

\bibitem[{Chen and Ferrara(2023)}]{chen2023tweets}
Chen, E.; and Ferrara, E. 2023.
\newblock Tweets in Time of Conflict: A Public Dataset Tracking the Twitter Discourse on the War Between Ukraine and Russia.
\newblock In \emph{ICWSM 2023 - 17th International AAAI Conference on Web and Social Media}. arXiv preprint arXiv:2203.07488.

\bibitem[{Chen et~al.(2022)Chen, Jiang, Chang, Muric, and Ferrara}]{chen2022charting}
Chen, E.; Jiang, J.; Chang, H.-C.~H.; Muric, G.; and Ferrara, E. 2022.
\newblock Charting the information and misinformation landscape to characterize misinfodemics on social media: COVID-19 infodemiology study at a planetary scale.
\newblock \emph{JMIR Infodemiology}, 2(1): e32378.

\bibitem[{Davis et~al.(2016)Davis, Ciampaglia, Aiello, Chung, Conover, Ferrara, Flammini, Fox, Gao, Gon{\c{c}}alves et~al.}]{davis2016osome}
Davis, C.~A.; Ciampaglia, G.~L.; Aiello, L.~M.; Chung, K.; Conover, M.~D.; Ferrara, E.; Flammini, A.; Fox, G.~C.; Gao, X.; Gon{\c{c}}alves, B.; et~al. 2016.
\newblock OSoMe: the IUNI observatory on social media.
\newblock \emph{PeerJ Computer Science}, 2: e87.

\bibitem[{Ferrara et~al.(2014)Ferrara, De~Meo, Catanese, and Fiumara}]{ferrara2014detecting}
Ferrara, E.; De~Meo, P.; Catanese, S.; and Fiumara, G. 2014.
\newblock Detecting criminal organizations in mobile phone networks.
\newblock \emph{Expert Systems with Applications}, 41(13): 5733--5750.

\bibitem[{Fraser(2016)}]{f:2016}
Fraser, C. 2016.
\newblock An analysis of the emerging role of social media in human trafficking: Examples from labour and human organ trading.
\newblock 15(2): 98--112.

\bibitem[{Hundman et~al.(2018)Hundman, Gowda, Kejriwal, and Boecking}]{hundman2018always}
Hundman, K.; Gowda, T.; Kejriwal, M.; and Boecking, B. 2018.
\newblock Always lurking: Understanding and mitigating bias in online human trafficking detection.
\newblock In \emph{Proceedings of the 2018 AAAI/ACM Conference on AI, Ethics, and Society}, 137--143.

\bibitem[{Kejriwal and Szekely(2017{\natexlab{a}})}]{kejriwal2017information}
Kejriwal, M.; and Szekely, P. 2017{\natexlab{a}}.
\newblock Information extraction in illicit web domains.
\newblock In \emph{Proceedings of the 26th international conference on world wide web}, 997--1006.

\bibitem[{Kejriwal and Szekely(2017{\natexlab{b}})}]{kejriwal2017knowledge}
Kejriwal, M.; and Szekely, P. 2017{\natexlab{b}}.
\newblock Knowledge graphs for social good: An entity-centric search engine for the human trafficking domain.
\newblock \emph{IEEE Transactions on Big Data}, 8(3): 592--606.

\bibitem[{Khan(2023)}]{k:2023}
Khan, M.~I. 2023.
\newblock Job Scams in LinkedIn Posts: How to Spot and Avoid Them.

\bibitem[{Laczko and Gramegna(2003)}]{lg:2003}
Laczko, F.; and Gramegna, M.~A. 2003.
\newblock Developing Better Indicators of Human Trafficking.
\newblock \emph{The Brown Journal of World Affairs}, 10(1): 179--194.

\bibitem[{Lee et~al.(2021)Lee, Vajiac, Kulshrestha, Levy, Park, Jones, Rabbany, and Faloutsos}]{lvklpjrf:2021}
Lee, M.-C.; Vajiac, C.; Kulshrestha, A.; Levy, S.; Park, N.; Jones, C.; Rabbany, R.; and Faloutsos, C. 2021.
\newblock INFOSHIELD: Generalizable Information-Theoretic Human-Trafficking Detection.
\newblock In \emph{2021 IEEE 37th International Conference on Data Engineering (ICDE)}, 1116--1127.

\bibitem[{Luceri et~al.(2024)Luceri, Pant{\`e}, Burghardt, and Ferrara}]{luceri2024unmasking}
Luceri, L.; Pant{\`e}, V.; Burghardt, K.; and Ferrara, E. 2024.
\newblock Unmasking the Web of Deceit: Uncovering Coordinated Activity to Expose Information Operations on Twitter.
\newblock In \emph{WWW’24}.

\bibitem[{Mart{\'\i}nez-Mesa et~al.(2014)Mart{\'\i}nez-Mesa, Gonz{\'a}lez-Chica, Bastos, Bonamigo, and Duquia}]{mgbbd:2014}
Mart{\'\i}nez-Mesa, J.; Gonz{\'a}lez-Chica, D.~A.; Bastos, J.~L.; Bonamigo, R.~R.; and Duquia, R.~P. 2014.
\newblock Sample size: how many participants do I need in my research?
\newblock \emph{Anais brasileiros de dermatologia}, 89: 609--615.

\bibitem[{{Office to Monitor and Combat Trafficking in Persons}(2023)}]{dos:2024}
{Office to Monitor and Combat Trafficking in Persons}. 2023.
\newblock Trafficking in person report.
\newblock In \emph{Department of State annual report}.

\bibitem[{Palmquist(2023)}]{p:2023}
Palmquist, K. 2023.
\newblock 17 Common Job Scams and How To Protect Yourself.
\newblock \emph{Indeed}.

\bibitem[{Rao et~al.(2021)Rao, Morstatter, Hu, Chen, Burghardt, Ferrara, and Lerman}]{rao2021political}
Rao, A.; Morstatter, F.; Hu, M.; Chen, E.; Burghardt, K.; Ferrara, E.; and Lerman, K. 2021.
\newblock Political partisanship and antiscience attitudes in online discussions about COVID-19: Twitter content analysis.
\newblock \emph{Journal of medical Internet research}, 23(6): e26692.

\bibitem[{Sapienza et~al.(2018)Sapienza, Ernala, Bessi, Lerman, and Ferrara}]{sapienza2018discover}
Sapienza, A.; Ernala, S.~K.; Bessi, A.; Lerman, K.; and Ferrara, E. 2018.
\newblock DISCOVER: Mining Online Chatter for Emerging Cyber Threats.
\newblock In \emph{Companion of the The Web Conference 2018}, 983--990. International World Wide Web Conferences Steering Committee.

\bibitem[{Szekely et~al.(2015)Szekely, Knoblock, Slepicka, Philpot, Singh, Yin, Kapoor, Natarajan, Marcu, Knight et~al.}]{szekely2015building}
Szekely, P.; Knoblock, C.~A.; Slepicka, J.; Philpot, A.; Singh, A.; Yin, C.; Kapoor, D.; Natarajan, P.; Marcu, D.; Knight, K.; et~al. 2015.
\newblock Building and using a knowledge graph to combat human trafficking.
\newblock In \emph{The Semantic Web-ISWC 2015: 14th International Semantic Web Conference, Bethlehem, PA, USA, October 11-15, 2015, Proceedings, Part II 14}, 205--221. Springer.

\bibitem[{Tavabi et~al.(2019)Tavabi, Bartley, Abeliuk, Soni, Ferrara, and Lerman}]{tavabi2019characterizing}
Tavabi, N.; Bartley, N.; Abeliuk, A.; Soni, S.; Ferrara, E.; and Lerman, K. 2019.
\newblock Characterizing Activity on the Deep and Dark Web.
\newblock In \emph{Companion Proceedings of the 2019 World Wide Web Conference}, 206--213.

\bibitem[{Tobey et~al.(2024)Tobey, Li, {\"O}zalt{\i}n, Mayorga, and Caltagirone}]{tl:2024}
Tobey, M.; Li, R.; {\"O}zalt{\i}n, O.~Y.; Mayorga, M.~E.; and Caltagirone, S. 2024.
\newblock Interpretable models for the automated detection of human trafficking in illicit massage businesses.
\newblock \emph{IISE Transactions}, 56(3): 311--324.

\bibitem[{Tong et~al.(2017)Tong, Zadeh, Jones, and Morency}]{tzjm:2017}
Tong, E.; Zadeh, A.; Jones, C.; and Morency, L.-P. 2017.
\newblock Combating Human Trafficking with Multimodal Deep Models.
\newblock In Barzilay, R.; and Kan, M.-Y., eds., \emph{Proceedings of the 55th Annual Meeting of the Association for Computational Linguistics (Volume 1: Long Papers)}, 1547--1556. Vancouver, Canada: Association for Computational Linguistics.

\bibitem[{UN(2003)}]{un:2003}
UN. 2003.
\newblock Report of the Ad Hoc Committee on the Elaboration of a Convention against Transnational Organized Crime on the work of its first to eleventh sessions.
\newblock \emph{United Nation}.

\bibitem[{UNODC(2024)}]{un:2024}
UNODC. 2024.
\newblock Global Report on Trafficking in Persons.
\newblock \emph{United Nation}.

\bibitem[{Whitney et~al.(2018)Whitney, Jennex, Elkins, and Frost}]{wjef:2018}
Whitney, J.; Jennex, M.; Elkins, A.; and Frost, E. 2018.
\newblock Don’t want to get caught? don’t say it: The use of emojis in online human sex trafficking ads.
\newblock \emph{Hawaii International Conference on System Sciences}.

\bibitem[{Zhou, Luceri, and Ferrara(2023)}]{zhou2023unveiling}
Zhou, S.; Luceri, L.; and Ferrara, E. 2023.
\newblock Unveiling the Dynamics of Censorship, COVID-19 Regulations, and Protest: An Empirical Study of Chinese Subreddit r/china\_irl.
\newblock In \emph{ICWSM 2023 Companion Proceedings}. arXiv preprint arXiv:2304.02800.

\bibitem[{Zhu, Li, and Jones(2019)}]{zlj:2019}
Zhu, J.; Li, L.; and Jones, C. 2019.
\newblock Identification and Detection of Human Trafficking Using Language Models.
\newblock In \emph{2019 European Intelligence and Security Informatics Conference (EISIC)}, 24--31.

\end{thebibliography}
\end{document}